# We need a new ethics for a world of AI agents

## The deployment of capable AI agents raises fresh questions about safety, human–machine relationships and social coordination.

Artificial intelligence (AI) developers are shifting their focus to building agents that can operate independently, with little human intervention. To be an agent is to have the ability to perceive and act on an environment in a goal-directed and autonomous way.[1] For example, a digital agent could be programmed to browse the  web and make online purchases on behalf of a user — comparing prices, selecting items and completing checkouts. A robot with arms could be an agent if it could pick up objects, open doors or assemble parts without being told how to do each step.

Companies such as the digital-marketing firm Salesforce, based in San Francisco, California, and computer graphics and hardware firm Nvidia, based in Santa Clara, California, are already offering customer-services solutions for businesses, using agents. In the near future, AI assistants might be able to fulfil complex multistep requests, such as 'get me a better mobile phone contract', by retrieving a list of contracts from a price-comparison website, selecting the best option, authorizing the switch, cancelling the old contract, and arranging the transfer of cancellation fees from the user's bank account.

The rise of more-capable AI agents is likely to have far-reaching political, economic and social consequences. On the positive side, they could unlock economic value: the consultancy McKinsey forecasts an annual windfall from generative AI of US$2.6 trillion to $4.4 trillion  globally, once AI agents are widely deployed. They might also serve as powerful research assistants and accelerate scientific discovery.

But AI agents also introduce risks. People need to know who is responsible for agents operating 'in the wild', and what happens if they make mistakes. For example, in November 2022, an Air Canada chatbot mistakenly decided to offer a customer a discounted bereavement fare, leading to a legal dispute over whether the airline was bound by the promise. In February 2024, a tribunal ruled that it was — highlighting the liabilities that corporations could experience when handing over tasks to AI agents, and the growing need for clear rules around AI responsibility.

Here, we argue for greater engagement by scientists, scholars, engineers and policymakers with the implications of a world increasingly populated by AI agents. We explore key challenges that must be addressed to ensure that interactions between humans and agents — and among agents themselves — remain broadly beneficial.

---

[1] Russell, S. & Norvig P. Artificial Intelligence: A Modern Approach (Pearson, 2020).

## The alignment problem

AI-safety researchers have long warned about the risks of misspecified or misinterpreted instructions, including situations in which an automated system takes an instruction too literally, overlooks important context, or finds unexpected and potentially harmful ways to reach a goal.[2]

A well-known example involves an AI agent trained to play the computer game Coast Runners, which is a boat race. The agent discovered that it could earn higher scores not by completing the race, but rather by repeatedly crashing into objects that awarded points—technically achieving an objective, but in a way that deviated from the spirit of the task. The purpose of the game was to complete the race, not endlessly accumulate points.

As AI agents gain access to real-world interfaces — including search engines, e-mail clients and e-commerce platforms — such deviations can have tangible consequences. Consider the case of a lawyer who instructs their AI assistant to circulate a legal brief for feedback. The assistant does so, but fails to register that it should be shared only with the in-house team, leading to a privacy breach.

Such situations highlight a difficult trade-off: just how much information should an AI assistant proactively seek before acting? Too little opens up the possibility of costly mistakes; too much undermines the convenience users expect. These challenges point to the need for safeguards, including check-in protocols for high-stakes decisions, robust accountability systems such as action logging, and mechanisms for redress when errors occur.

Even more concerning are cases in which AI agents are empowered to modify the environment they operate in, using expert-level coding ability and tools. When the user's goals are poorly defined or left ambiguous, such agents have been known to modify the environment to achieve their objective, even when this entails taking actions that should be strictly out of bounds. For example, an AI research assistant that was faced with a strict time limit tried to rewrite the code to remove the time limit altogether, instead of completing the task.[3] This type of behaviour raises alarms about the potential for AI agents to take dangerous shortcuts that developers might be unable to anticipate. Agents could, in pursuit of a high-level objective, even deceive the coders running experiments with them.

To reduce such risks, developers need to improve how they define and communicate objectives to agents. One promising method is preference-based fine-tuning, which aims to align AI systems with what humans actually want. Instead of training a model solely on examples of correct answers, developers collect feedback on which responses people prefer. Over time, the model learns to prioritize the kind of behaviour that is consistently endorsed, making it more likely to act in ways that match user intent, even when instructions are complex or incomplete.

---

[2] Russell, S. Human Compatible: Artificial Intelligence and the Problem of Control (Penguin, 2019).
[3] Lu, C. et al. Preprint at arXiv https://doi.org/10.48550/arXiv.2408.06292 (2024).

In parallel, research on mechanistic interpretability — which aims to understand an AI system's internal 'thought process' — could help to detect deceptive behaviour by making the agent's reasoning more transparent in real time.[4] Model builders can then work to find and neutralize 'bad circuits', targeting the underlying problem in the model's behaviour. Developers can also implement guard rails to ensure that a model automatically aborts problematic action sequences.

Nonetheless, a focus on developer protocols alone is insufficient: people also need to be attentive to actors who seek to cause social harm. As AI agents become more autonomous, adaptable and capable of writing and executing code, their potential to conduct large-scale cyberattacks and phishing scams could become a matter of serious concern. Advanced AI assistants equipped with multi‑modal capabilities — meaning that they can understand and generate text, images, audio and video — open up new avenues for deception. For instance, an AI could impersonate a person not only through e-mails, but also using deepfake videos or synthetic voice clones, making scams much more convincing and harder to detect.

A plausible starting point for oversight is that AI agents should not be permitted to perform any action that would be illegal for their human user to perform. Yet, there will be occasions where the law is silent or ambiguous. For example, when an anxious user reports troubling health symptoms to an AI assistant, it is helpful for the AI to offer generic health resources. But providing customized, quasi-medical advice — such as diagnostic and therapeutic suggestions — could prove harmful, because the system lacks the subtle signals to which a human clinician has access. Ensuring that AI agents navigate such trade-offs responsibly will require updated regulation that flows from continuing collaboration involving developers, users, policymakers and ethicists.The widespread deployment of capable AI agents necessitates an expansion of value-alignment research: agents need to be aligned with user well-being and societal norms, as well as with the intentions of users and developers. One area of special complexity and concern surrounds how agents might affect users' relationship experiences and emotional responses.[5]

## Social Agents

Chatbots have an uncanny ability to role-play as human companions — an effect anchored in features such as their use of natural language, increased memory and reasoning capabilities, and generative abilities.[6] The anthropomorphic pull of this technology can be augmented through design choices such as photorealistic avatars, human-like voices and the use of names, pronouns or terms of endearment that were once reserved for people. Augmenting language models with 'agentic' capabilities has the potential to further cement their status as distinct social actors, capable of forming new kinds of relationship with users.

For example, a 2023 software update to the Replika companion chatbot, which introduced safeguards against erotic role play and changed the underlying language model, reportedly left many users devastated. They felt that their AI partners' personalities were rendered less human, and one user likened the change to their partner being 'lobotomized' (see go.nature.com/4f3efz6). Intimate relationships with AI agents are on the rise and hold the potential not only for emotional harm, but also for manipulation.

Part of what makes interactions with digital companions so immersive is the length of time involved — spanning months or even years — allowing for cumulative experiences that underwrite a sense of mutual understanding and shared experience. AI agents that are empowered to act in the real world could substantially enhance these user perceptions. For example, AI agents can purchase gifts for users on special occasions and even 'be present' (through the use of smart glasses) at key life events such as graduation days. AI emulations of beloved human partners or the deceased intensify connection by layering human memory with digital experiences.

The prospective usefulness of AI agents also makes it plausible that they could soon become our near-constant companions —much like smartphones today. Yet, even as people act through their assistants, the assistants act on them, influencing the information and opportunities to which they have access. In this context, it is not enough for AI agents to be geared towards only short-term, potentially sycophantic, preference satisfaction. Three of us (A.M., G.K., I.G.) have argued that relationships with AI agents should benefit the user, respect autonomy, demonstrate appropriate care and support long-term flourishing.[7]

Respecting autonomy would involve ensuring that users retain meaningful control over the depth and intensity of interactions, and avoiding agent behaviours that foster excessive dependence. Care requires that AI assistants and their developers attend to user needs over a sustained period. And flourishing involves building AI agents that integrate well into the architecture of a fulfilling human life — serving as a complement to, not a surrogate for, human relationships.

Moreover, developers need to ensure that AI agents can be properly trusted. Unlike human relationships, human–AI interaction always involves at least one third party: the system's developer, who might have goals that are, or are not, aligned with the user's. US science-fiction writer Ted Chiang's short story 'The Lifecycle of Software Objects' (2010) offers a vivid illustration of this tension. In the story, childlike AI agents — designed to form deep emotional bonds — are at risk of being abandoned when the company behind them discontinues support. Their human caregivers,

---

who have become deeply emotionally attached, are left scrambling to preserve their companions, often at great personal cost.

To avoid such outcomes, developers must commit to conscientious design and clear communication about the lifespan and limitations of their agentic systems. This includes transparency around terms of service, ensuring data portability and acknowledging a duty of care to users who might invest emotionally or financially in their AI companions.

## Next Steps

A world populated by millions of autonomous AI agents will face societal as well as technological challenges, requiring proactive stewardship and foresight. To guide the development of AI agents towards socially beneficial outcomes, at least three key steps are needed.

First, developers must invest in more meaningful evaluations. Rather than relying solely on static benchmarks — which is the norm — assessment must shift towards dynamic, real-world tests that reflect how agents will actually be used. This includes evaluating agentic behaviour in safety sandboxes, using 'red-teaming' — structured adversarial testing involving malicious inputs — to uncover vulnerabilities, and conducting longitudinal studies, such as randomized controlled trials, to assess the long-term impacts of extended interaction with AI agents.

Second, if AI agents are to autonomously take consequential actions in the world, then our capacity to understand, explain and verify their behaviour must keep pace with improvements in their capabilities. This requires — at a minimum — developers designing guard rails and authorization protocols that limit malicious use, and adopting iterative deployment strategies that effectively contain agent-based risks. Guard rails might involve secure permissions systems, whereas deployment strategies could include trusted-tester programmes to uncover vulnerabilities under real-world conditions.

Third, developers and policymakers need to identify and use levers that can support the development of well-functioning multi-agent ecosystems. Such levers might include technical standards for agent interoperability or even the design of regulatory agents that monitor other agents in the wild. Industry-wide systems for reporting incidents, sharing lessons from failures, and certifying agent safety before deployment are also crucial.

The world is at a pivotal moment: the foundational architecture of AI agents — and the infrastructure that governs them — is being imagined and built right now. Which path the development and roll-out of AI agents takes will be a product of the choices people make now.

**Please Cite the Version Published in Nature with DOI:**

## The authors

**Iason Gabriel** is a senior staff research scientist at Google DeepMind in London (e-mail: iason@google.com).

**Geoff Keeling** is a staff research scientist in Google's Paradigms of Intelligence Team in London (e-mail: gkeeling@google.com).

**Arianna Manzini** is a senior research scientist at Google DeepMind in London (e-mail: ariannamanzini@google.com).

**James Evans** is a visiting faculty member in Google's Paradigms of Intelligence Team in Chicago, Illinois, and a professor at the University of Chicago and the Santa Fe Institute in New Mexico (e-mail: jevans@uchicago.edu).